\documentclass[aps,floatfix,twocolumn,a4paper,showpacs, nofootinbib,
superscriptaddress,10pt]{revtex4}
\usepackage{graphicx,float}\usepackage{graphicx,float}
\usepackage[all]{xy}
\usepackage{amsmath,upgreek}
\usepackage{amssymb}
\usepackage{color}
\usepackage{epsfig}		
\usepackage{graphicx,epstopdf}
\usepackage{subfigure}
\usepackage{pdfpages}
\usepackage[colorlinks,hyperindex]{hyperref}

\definecolor{green1}{RGB}{0,128,0} 
\hypersetup{hidelinks,backref=true,pagebackref=true,hyperindex=true,colorlinks=true,breaklinks=true,urlcolor= blue}
\hypersetup{%
  colorlinks = true,
  linkcolor  = blue,
  citecolor = green1,
}
\usepackage{bookmark,textgreek}
\usepackage{hyperref,color,xcolor}
\hypersetup{hidelinks,hyperindex=true,colorlinks=true,breaklinks=true,urlcolor= blue}
\hypersetup{%
  colorlinks = true,
  linkcolor  = blue
}

\newcommand{\bes}{\begin{subequations}}
\newcommand{\ees}{\end{subequations}}
\def\ben{\begin{eqnarray}}
\def\een{\end{eqnarray}}
\def\be{\begin{equation}}
\def\ee{\end{equation}}

\begin{document}

\title{Pion family in AdS/QCD: the next generation from  
configurational entropy }

\author{Luiz F.  Ferreira  }\email{luizfaulhaber@if.ufrj.br}
\affiliation{Federal University of ABC, Center of Mathematics,  Santo Andr\'e, Brazil}
\affiliation{Federal University of ABC, Center of Physics,  Santo Andr\'e, Brazil.}
\author{R. da Rocha}
\email{roldao.rocha@ufabc.edu.br}
\affiliation{Federal University of ABC, Center of Mathematics, Santo Andr\'e, Brazil}\email{roldao.rocha@ufabc.edu.br}

\begin{abstract}
The two flavour AdS/QCD, with chiral and gluon condensates, sets in the description of the pion family and 
  its mass spectra. Using gravity-dilaton-gluon backgrounds, entropic Regge-like trajectories for the pion family are then derived. They relate the $\pi$ mesons underlying configurational entropy to both the pions excitation wave mode number and 
the pions experimental mass spectra, yielding a reliable prediction for the mass spectra of higher excitation pion modes, to be 
experimentally detected. 
 \end{abstract}
\maketitle

\section{Introduction}
The configurational entropy (CE) apparatus consists in measuring the shape complexity of any localized system \cite{Gleiser:2011di,Gleiser:2012tu}.  The information encodement in random processes is  represented by the CE, that thus implements  the  compression of information in the wave mode configurations endowing any system in Nature \cite{Gleiser:2011di,Gleiser:2012tu}. The CE setup has been shown to be an advantageous instrument to probe vast aspects of QCD, in the holographic AdS/QCD correspondence, both from theoretical and phenomenological aspects. 
Elementary particles and their excitations, and some of their features, in QCD, were shown to correspond to critical points of the CE \cite{Bernardini:2016hvx,Braga:2017fsb,Barbosa-Cendejas:2018mng,Braga:2018fyc,Bernardini:2016qit,daSilva:2017jay,Ma:2018wtw}. Among other fields, mainly mesonic states and glueballs play a prominent role in the informational aspects of AdS/QCD.
The phenomenological dominance and abundance of certain excitation levels of mesonic states -- in particular 
light-favour mesons and quarkonia --  and glueballs is an immediate implication of encoding and compressing information into the 
Fourier wave modes that underlie and constitute such states. Quantum states whose CE is bigger were shown to be less observable in experiments. Phenomenological aspects of the CE in AdS/QCD was studied in Refs.  \cite{Bernardini:2016hvx,Barbosa-Cendejas:2018mng} for light-flavour mesons, in Refs. \cite{Braga:2017fsb,Braga:2018fyc} for bottomonium and charmonium, both in the QFT and in the finite temperature setups, and in Ref. \cite{Bernardini:2016qit} for scalar glueballs. Other aspects of the CE in QCD were paved in Refs. \cite{Karapetyan:2016fai,Karapetyan:2017edu,Karapetyan:2018oye,Karapetyan:2018yhm,Karapetyan:2019fst} and in the standard model as well, in Refs. \cite{Alves:2014ksa,Alves:2017ljt}.  Informational entropic Regge-like trajectories of the $a_1$, $f_0$ and $\rho$  light-flavour mesons were also introduced in Ref.  
 \cite{Bernardini:2018uuy}, in the context of the CE.  In addition, the CE is a sharp tool that can also probe
  phase transitions, identified to CE critical points,  in diverse systems in Physics \cite{Gleiser:2014ipa,Gleiser:2018kbq,Sowinski:2017hdw,Sowinski:2015cfa,Gleiser:2013mga,Gleiser:2015rwa,Casadio:2016aum,Braga:2016wzx,roldao,Lee:2018zmp,Bazeia:2018uyg,Correa:2016pgr}. A more precise review of  information theory, in both its classical and quantum aspects, can be checked in Ref. \cite{Witten:2018zva}.

In the last  years, the AdS/QCD models has been used to study  many non-perturbative aspects of QCD,  such as  confinement \cite{Karch:2006pv,Zhang:2010tk,Colangelo:2011sr,Colangelo:2018mrt}, hadronic mass spectra \cite{Li:2013oda,Brodsky:2014yha} and chiral symmetry breaking \cite{rold,Gkk}. The AdS/QCD setup, inspired by the gauge/gravity duality \cite{Maldacena:1997re},  considers a similar type of duality, where the conformal invariance is broken after introducing an energy parameter in the AdS bulk. The simplest one AdS/QCD model is the hard wall one \cite{Polchinski:2001tt,BoschiFilho:2002ta,BoschiFilho:2002vd} and consists in placing a hard geometrical cut-off into the AdS space. Another AdS/QCD model is represented by the soft wall technique \cite{Karch:2006pv}, reproducing  a very important feature, the so-called linear Regge behaviour. In this case, the background involves the AdS space and a scalar field, the dilaton, that  effectively acts as a smooth infrared cut-off.

Among the mesons that can be  investigated  in the AdS/QCD models, the pseudoscalar $\pi$ meson, the pion, occupies a prominent position.  There are many studies of the properties in holographic approaches \cite{EKSS2005,Grigoryan:2007wn,Brodsky:2007hb,Kelley:2010mu,sui1,Li:2013oda,Ballon-Bayona:2014oma,Lv:2018wfq}. In particular, the approach of Ref. \cite{Li:2013oda}  constructs a dynamical hQCD model for the two flavor theory, in the graviton-dilaton background, that incorporates  the Regge behaviour of the hadron spectra, the linear quark potential and the chiral symmetry breaking. It was found in this model that the lowest pion state has a mass around $140 $ MeV, which can be regarded as the Nambu-Goldstone bosons due to the chiral symmetry breaking. The mass spectra of the pion agrees well with the experimental data \cite{pdg1}. 

Endowed with the tools of the CE and two flavour AdS/QCD, with chiral and gluon condensates backgrounds coupled to gravity, 
the configurational entropic Regge-like trajectories for the pion family shall be derived and scrutinized. There are two types of Regge-like trajectories to be studied. The first one relates the pion family CE to the pions excitation wave mode number, whereas the second one associates the pion family CE to the pions experimental mass spectra. Analyzing both types of Regge-like trajectories can, then, derive a reliable prediction for the mass spectra of higher excitation modes of $\pi$ mesonic  states. 
This paper is organized as follows: Sect. \ref{sec2} introduces the two flavour AdS/QCD soft wall model and the pion family EOM and main features, in this context. The mass spectra of the pion family is then depicted, showing the standard Regge trajectory. Sect. \ref{iii} is devoted to study the gluon and chiral condensates backgrounds couples with gravity, for two dilatonic models.
The CE is then calculated for the $\pi$ meson family, first with respect to the $n$ excitation mode number for the $\pi_n$ modes in the pion family. Besides, the CE is also computed as a function of the pion family mass  spectra, generating then a second type of non-linear Regge-trajectory. Therefore the two types of configurational entropic Regge trajectories are interpolated, relating the CE to both $n$ and the pion mass spectra. Hence, the mass spectra of higher excitation $\pi_n$ modes is inferred. It  provides a reliable phenomenological prediction of the mass spectra of the next generation of higher excitations in the pion family.  Sect. \ref{iv} 
closes the paper with discussions, conclusions and some perspectives.

\section{Two flavour AdS/QCD and the pion family }
\label{sec2}
\begin{figure}[H]
\centering
\includegraphics[width=8.5cm]{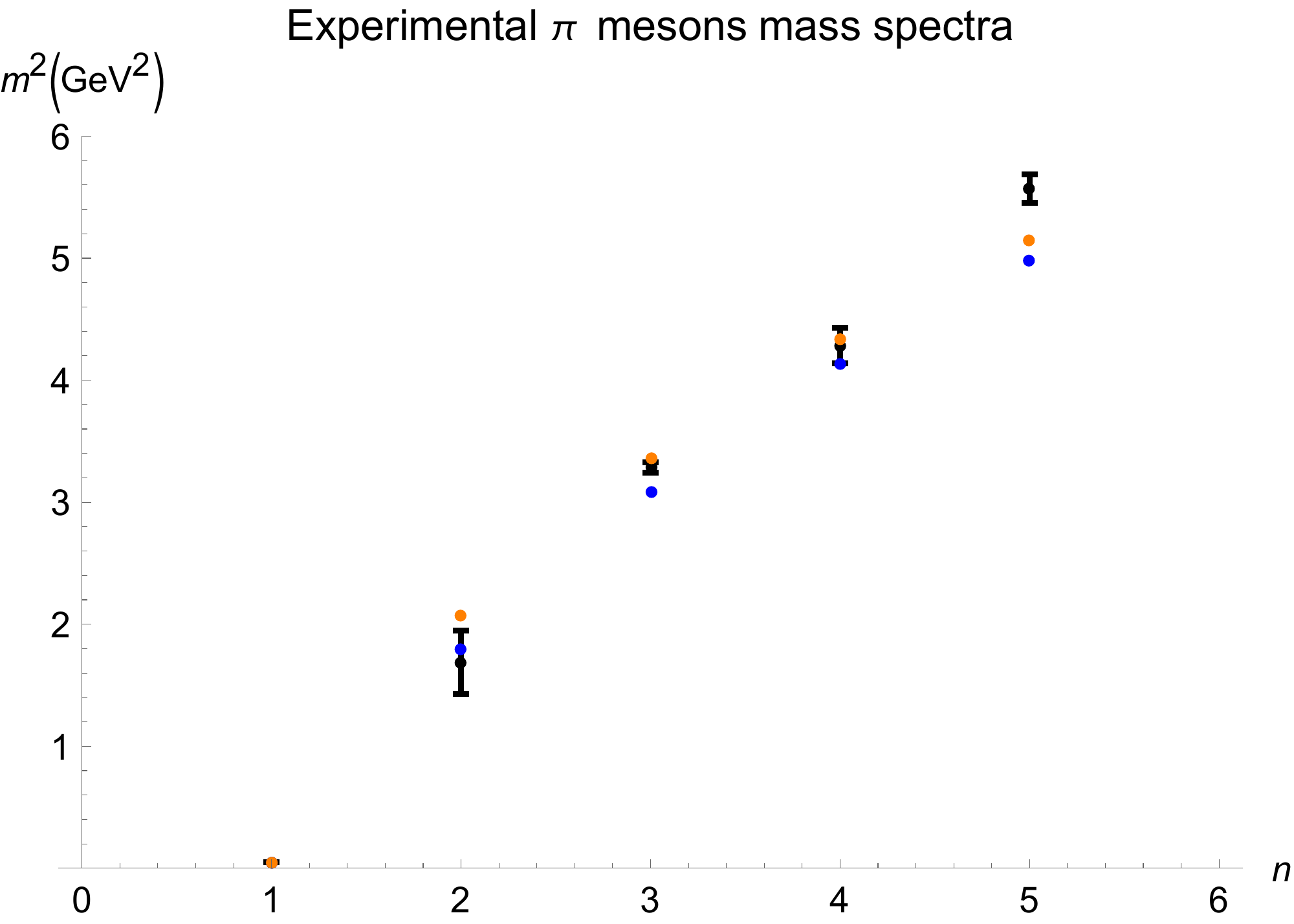}
\caption{\textcolor{black}{\textcolor{black}{Pion mass spectra as a function of the 
$n$ excitation number:  experimental values (black points) and the AdS/QCD model
predictions for the background modification given by Eqs. (\ref{quadraticd}) (blue points) and (\ref{tanh}) (orange points)}. The ${\mathtt{\pi}}_0$, ${\mathtt{\pi}}_\pm$, $\pi(1300)$, $\pi(1800)$ pseudoscalar  mesons, constitute already confirmed
 mesons in PDG. The last two $\pi(2070)$ and  
$\pi(2360)$ states are omitted from the summary table in PDG \cite{pdg1}}.}
\label{data1}
\end{figure}

Meson families were comprehensively described in AdS/QCD, both in the hard and soft wall models  \cite{Karch:2006pv,Braga:2015jca,Braga:2018zlu,Colangelo:2008us,Gkk,sui1,Afonin:2012jn}. Dynamical  holographic models were also approached in Refs. \cite{Rougemont:2017tlu,dePaula,dePaula:2009za,Barbosa-Cendejas:2018mng,BallonBayona:2007qr}. 
The so-called Regge trajectories, relating the excitation number of a light-flavour mesonic state and its mass, 
 were originally derived in Ref. \cite{Karch:2006pv}. Aiming to focus our approach into the pion family in the AdS/QCD, 
  the  bulk ${\rm AdS}_5$ metric is an important ingredient, reading  
\begin{equation}\label{bulkm}
ds^{2}=g_{MN}d{\rm x}^{M}d{\rm x}^{N}=e^{2 A({\rm z})}\!\left( \upeta_{\mu\nu}d{\rm x}^{\mu}d{\rm x}^{\nu}+d{\rm z}^2\right),
\end{equation}
where the warp factor $e^{A({\rm z})}=-{\rm z}/\ell$, and the $ \upeta_{\mu\nu}$ represents the boundary metric, being $\ell$ the ${\rm AdS}_5$  curvature radius. 
Mesonic excitations can be described by $\mathbb{X}({\rm z})$ fields in the ${\rm AdS}_5$ bulk. These fields represent the dual description  of the $q\bar{q}$ operator, with mass $m_\mathbb{X}$, that is ruled by the following action \cite{Karch:2006pv}
\begin{eqnarray}
 S=-\int \, e^{-\upphi({\rm z})} \sqrt{-g}\,Tr\,\mathbb{L}\,d^4{\rm x}\, d{\rm z}, \label{softw}
 \end{eqnarray} 
where  
\begin{eqnarray}\label{lagran}
\!\!\!\!\!\mathbb{L}\sim D^M\mathbb{X}D_M\mathbb{X}+m_\mathbb{X}^2 \mathbb{X}^M\mathbb{X}_M
 +\mathbb{F}_L^2+\mathbb{F}_R^2,
 \end{eqnarray}  
 being the $\mathbb{A}_L^{m}$ and
$\mathbb{A}_R^{m}$ gauge fields that drive the SU(2)$_R\times$ SU(2)$_L$ chiral flavour symmetry of QCD. The Yang--Mills field strengths are given by  
\begin{eqnarray}
\mathbb{F}_{R,L}^{MN}&=&\partial^{[M}{A_{R,L}^{N]}}-i[A_{R,L}^{M},A_{R,L}^{N}],\end{eqnarray}
where $A_{R,L}^{M}= A_{R,L}^{Mc} f_c$, being  $\{f_c\}$ ({\small{$c=1,2,3$}}) SU(2) generators. The covariant derivative is explicitly given by 
$D^M \mathbb{X}=\partial^M \mathbb{X}-i \mathbb{A}_L^M \mathbb{X}+i\mathbb{X}, \mathbb{A}_R^M.
$ 
The $\mathbb{X}({\rm z})$ field  is constructed upon the a pseudoscalar field, P, and a scalar field, S, as \cite{Li:2013oda}
\begin{equation}
\mathbb{X}({\rm z}) = \exp\left(i{\rm P}^{c}t^{c}\right)\;\left({\upxi({\rm z})}+{\rm S}\right),
\label{scalarfield}
\end{equation}
where $\upxi({\rm z})$ is a vacuum expectation
value that breaks chiral symmetry \cite{Li:2013oda}. 
To describe the vector and axial vector meson, the left, $\mathbb{A}_L$, and right, $\mathbb{A}_R$, gauge
fields can be split into vector, V, and axial vector, A,  fields, as  \cite{Li:2013oda} 
\begin{subequations}
\begin{eqnarray}
{\rm A}^M&=&\frac12(\mathbb{A}^M_R-\mathbb{A}^M_L),\label{a12} \\
V^M&=&\frac12(\mathbb{A}^M_R+\mathbb{A}^M_L)\label{vv12}
\end{eqnarray}
\end{subequations}
yielding the respective gauge field strengths, 
\begin{eqnarray}
F_{V}^{MN}&=&\partial^{[M}{V^{N]}}+{i}[V^{N},V^{M}],\\
F_{{\rm A}}^{MN}&=&\partial^{[M}{{\rm A}^{N]}}+{i}\left[{\rm A}^{N},{\rm A}^{M}\right].
\end{eqnarray}
With respect to the vector $V$ and axial vector ${\rm A}$ fields, the soft wall Lagrangian (\ref{lagran})
reads 
\begin{equation}
\!\!\!\!\!\!\!\mathbb{L}\sim{\rm D}^M\mathbb{X}{\rm D}_M\mathbb{X}+m_\mathbb{X}^2 \mathbb{X}^M\mathbb{X}_M
 +2\left(F_{{\rm A}}^{2}+F_{V}^{2}\right),
\label{softwvecax}
\end{equation}
for ${\rm D}^M\mathbb{X}=\partial_M\mathbb{X}+i\left[\mathbb{X},V_M\right]-i\left\{A_M,\mathbb{X}\right\}$.  
The EOM for the $\upxi({\rm z})$ field then reads 
\begin{eqnarray}
\!\!\!\!\!\!\!\!\left[\frac{d^2}{d{\rm z}^2}\!-\!(\upphi'({\rm z})-3 {\rm A}'({\rm z}))\frac{d}{d{\rm z}}\!-\! m_\mathbb{X}^2({\rm z}) e^{2 A({\rm z})}\right]\upxi({\rm z})\!=\!0.
\label{vevo}
\end{eqnarray}
The analytical expression, as well as the plots of the $\upxi({\rm z})$ field can be found in Refs. \cite{Li:2013oda,Bernardini:2018uuy}. 
Let one denotes by ${\mathtt{\pi}}_n$ the functions that describe the pions,
and by ${\tt \varphi}_n$ those ones that represent the $\phi$  mesons.  It is worth to emphasize that, as in Ref. \cite{Bernardini:2018uuy}, the ${\mathtt{\pi}}_n({\rm z})$ modes are regarded as the quantum mechanical wave functions, $\uppsi_n({\rm z})$, by ${\mathtt{\pi}}_n({\rm z})=z^{1/2}\,\exp({\rm z}^2/2)\,\uppsi_n({\rm z})$. Standardly the $n=1$ index often alludes to the ground state.
Therefore, the functional form for $\uppsi_n({\rm z})$ reads $
\uppsi_n({\rm z}) \sim e^{-{\rm z}^2/2}\,{\rm z}^{m+1/2}L^m_{n-1}({\rm z}^2)$, 
that shall be used in the forthcoming calculations, to correspond the $n=1$ level for the standard ground state, taking the same indexing as the Hydrogen radial quantum number problem.
The EOMs driving the  pion and $\phi$ meson families read
\begin{eqnarray}
\!\!\!\!\!\!\!\!\!\!\!\!\!\!\!\left(-\frac{d^2}{d{\rm z}^2}\!+\!V_{\pi}({\rm z})\right)\pi_n({\rm z})\!\!&\!\!=\!\!&\!\!m_n^2 ({\mathtt{\pi}}_n({\rm z})\!-\!e^A\upxi\upvarphi_n({\rm z})),\label{motionpi} \\
\!\!\!\!\!\!\!\!\!\!\!\!\!\!\left(-\frac{d^2}{d{\rm z}^2}+V_{\varphi}({\rm z})\right)\upvarphi_n({\rm z})\!\!&\!=\!&\! e^A\upxi(\pi_n({\rm z})-e^A\upxi\upvarphi_n({\rm z})),\label{motionphi}
\end{eqnarray}
 where the Schr\"odinger-like potentials are respectively given by 
\begin{eqnarray}
V_{\pi}({\rm z})&=& \frac{{\rm A}''}{2}-\frac{\upphi{''}}{2}+\frac{\upxi{''}}{\upxi}-(\log\upxi)'^{2} \nonumber\\&&\qquad\qquad+\frac{1}{4}(\upphi{'}-3{\rm A}{'}-2(\log\upxi){'})^2.\label{v12}\\
V_{\varphi}({\rm z})&=&-\frac{\upphi{''}-{\rm A}{''}}{2}+\frac{(\upphi{'}-{\rm A}{'})^2}{4}.
 \label{v11}
\end{eqnarray}
Endowed with the scalar potential (\ref{v12}), the mass spectra of
 the  pion family can be read off Eq. (\ref{motionpi}). 
 As the only element in the $\phi_n$ meson family that was experimentally observed is the $\phi(1020)$, whereas the $\phi(1680)$  
 is not yet included in the PDG summary table, the analysis concerning the pion family shall be emphasized in what follows.
 In fact, experimental data regarding the $\phi_n$ family is very scarce.   
The experimental mass spectra of the pion family is depicted in Fig. \ref{data1}. 
The ${\mathtt{\pi}}_0,\pi_\pm$, $\pi(1300)$, $\pi(1800)$ pseudoscalar  mesons have been already experimentally confirmed
 meson particles in PDG. On the other hand, the $\pi(2070)$ and  
$\pi(2360)$ excitations are omitted from the summary table in PDG \cite{pdg1}. 
The Regge trajectory for the pion meson family can be extrapolated in Fig. \ref{data1}. 
%
\section{Configurational entropic Regge trajectories in two flavour AdS/QCD}
\label{iii}
The  AdS/QCD setup can be then employed to derive configurational entropic Regge trajectories for the pion family.
Based upon a two flavour soft wall model, with gluon and chiral condensates, coupled to gravity with a dilaton \cite{Colangelo:2011sr}, 
informational Regge trajectories were studied to the $a_1$, $f_0$ and $\rho$ meson families in such a setup \cite{Bernardini:2018uuy}.
The following dilatons were introduced in Refs.  \cite{Li:2013oda,Bernardini:2018uuy} to model mesons and glueballs, 
\begin{eqnarray}
\upphi_1({\rm z})&=&\upmu_{\rm G}^2{\rm z}^2, \label{quadraticd} \\
 \upphi_2({\rm z})&=&\upmu_{\rm G}^2{\rm z}^2\tanh\left({\upmu_{{\rm G}^2}^4{\rm z}^2}/{\upmu_{\rm G}^2}\right).
\label{tanh}
\end{eqnarray}
and shall be employed, respectively as the prototypical dilaton in the soft wall AdS/QCD \cite{Karch:2006pv}, Eq. (\ref{quadraticd}), and its deformation, Eq. (\ref{tanh}). The deformed dilaton in the UV limit yields the quadratic dilaton. The holographic gluon condensate is dual to the quadratic dilaton (\ref{quadraticd}) and has $\upmu_{\rm G}$ energy scale when it corresponds to a  dimension-2 system, whereas it has $\upmu_{\rm G}^{2}$ energy scale when describing a dimension-4 dual system \cite{Gkk,Afonin:2012jn}. 
A graviton-gluon-dilaton action in AdS can be given by \cite{Li:2013oda}, 
\begin{eqnarray}\label{alll}
\!\!\!\!\!\!\!\!S\!=\!\kappa_5^2 \int \, \sqrt{-g}e^{-2\upphi}\Big\{ \left[R+4\partial^M\upphi
\partial_M \upphi - 4V_g(\upphi) \right.\nonumber\\\left.
 \!\!\!\!\!\!\!\!\!\!\!\!\!\!\hspace*{-2cm}- 16\lambda e^{-\upphi}\!\left(\partial^M\upxi \partial _M \upxi
\!+\! V(\upphi,\upxi)\right)\right]\!\Bigg\}\,d^5 x,
\end{eqnarray}
where $\lambda$ denotes a general coupling, and 
$V_g$ denotes the gluon system potential. Ref. \cite{Li:2013oda} studied a heavy quark potential in the background given by Eq. (\ref{alll}), deriving the physical 
effective potential $V(\upphi,\upxi)\approx \upxi^2\upphi^2$. 
For both the $\upphi_1({\rm z})$ and $\upphi_2({\rm z})$, respectively in Eqs. (\ref{quadraticd}) and (\ref{tanh}), the parameters 
$\upmu_{{\rm G}^2}=\upmu_{\rm G}\approxeq 0.431$ were adopted in Refs. \cite{Li:2013oda,Bernardini:2018uuy}, in full compliance to data 
from experiments in PDG. 
Numerical analysis of the EOMs derived from (\ref{alll}), in Ref. \cite{Li:2013oda}, yields the solutions for $\upxi({\rm z})$,  for  both the dilatonic backgrounds.
  \textcolor{black}{The first column in Table \ref{scalarmasses} replicates the mass spectra in the PDG 2018 for   
${\mathtt{\pi}}_1=\{{\mathtt{\pi}}_\pm,  
{\mathtt{\pi}}_0\}$, ${\mathtt{\pi}}_2=\pi(1300)$, ${\mathtt{\pi}}_3=\pi(1800)$, as well as for ${\mathtt{\pi}}_4=\pi(2070)$, ${\mathtt{\pi}}_5=\pi(2360)$ that are still left out the summary table in PDG (few events registered \cite{pdg1}).}
\begin{table}[h]
\begin{center}--------- pseudoscalar pion mass spectra (MeV) ---------\medbreak
\begin{tabular}{||c|c||c|c||}
\hline\hline
        $n$ & ~Experimental         & mass$_{\upphi_1({\rm z})}$&  mass$_{\upphi_2({\rm z})}$  \\\hline\hline
\hline
         \textcolor{black}{1} & $139.57018\pm0.00035$    &139.3     &139.6              \\\hline
       \textcolor{black}{1} &   $134.9766 \pm 0.0006$&139.3     &139.6\\\hline
         \textcolor{black}{2}& $1300 \pm 100$           &1343      &1505             \\\hline
         \textcolor{black}{3} & $1816 \pm 14 $         &1755               &1832   \\\hline
         \textcolor{black}{4*} & $2070$           &2006                 &2059   \\\hline
         \textcolor{black}{5*} & $2360$           &2203     &2247                \\
\hline\hline
\end{tabular}
\caption{Mass spectra for the pseudoscalar pion family, in the $\upphi_2({\rm z})={\rm z}^2\tanh\left({\upmu_{{\rm G}^2}^4{\rm z}^2}/{\upmu_{\rm G}^2}\right)$  dilaton, for the $\pi_0$, $\pi(1300)$, $\pi(1800)$, $\pi(2070)$, 
$\pi(2360)$ mesons. The modes indicated with asterisk are not established particles and
therefore are omitted from the summary table in PDG. }
 \label{scalarmasses}
\end{center}
\end{table}

Table  \ref{scalarmasses} shows the pseudoscalar pion family, identifying the ${\mathtt{\pi}}_n$ eigenfunctions  in Eq. (\ref{motionphi}), as ${\mathtt{\pi}}_1=\{{\mathtt{\pi}}_\pm,  
{\mathtt{\pi}}_0\}$, ${\mathtt{\pi}}_2=\pi(1300)$, ${\mathtt{\pi}}_3=\pi(1800)$, ${\mathtt{\pi}}_4=\pi(2070)$, ${\mathtt{\pi}}_5=\pi(2360).$
 whereas the other ones have not been experimentally confirmed states yet \cite{pdg1}. Besides, the pseudoscalar sector can be implemented by considering the following action, 
pion and $\phi$ meson wavefunctions:
\begin{eqnarray}
S_{\pi}^{(2)} &=&
-\frac{1}{3L^3}\int d^5x
 e^{-\upphi}\sqrt{g}(\upxi^2\partial^{\rm z}\pi\partial_{\rm z}\pi
  \nonumber \\
 \!\!\!\!\!&&\!\!\!\!+\upxi^2\partial^\mu(\varphi\!-\!\pi)\partial_\mu(\varphi\!-\!\pi)\!+\!{L^2}\partial^{\rm z}\partial^\mu\varphi\partial_{\rm z}\partial_\mu\varphi).
\end{eqnarray}
It is observed that in the graviton-dilaton-scalar system, the lowest pseudoscalar state has
a mass around $140 {\rm MeV}$, which can be regarded as the Nambu-Goldstone bosons due
to the chiral symmetry breaking. The higher excitations yields a Regge trajectory, in full compliance to experimental data. \textcolor{black}{It is worth to mention that for both dilatonic field backgrounds (\ref{quadraticd}, \ref{tanh}) the so-called Gell-Mann--Oakes--Renner relation, stating that the square of the pion mass is proportional to the product of a sum of quark masses and the quark condensate, is satisfied, yielding the production of massless pions in the $m_q = 0$ limit, where $m_q$ denotes the quark mass \cite{EKSS2005,Grigoryan:2007wn}.}

\begin{figure}[H]
\centering
\includegraphics[width=8.3cm]{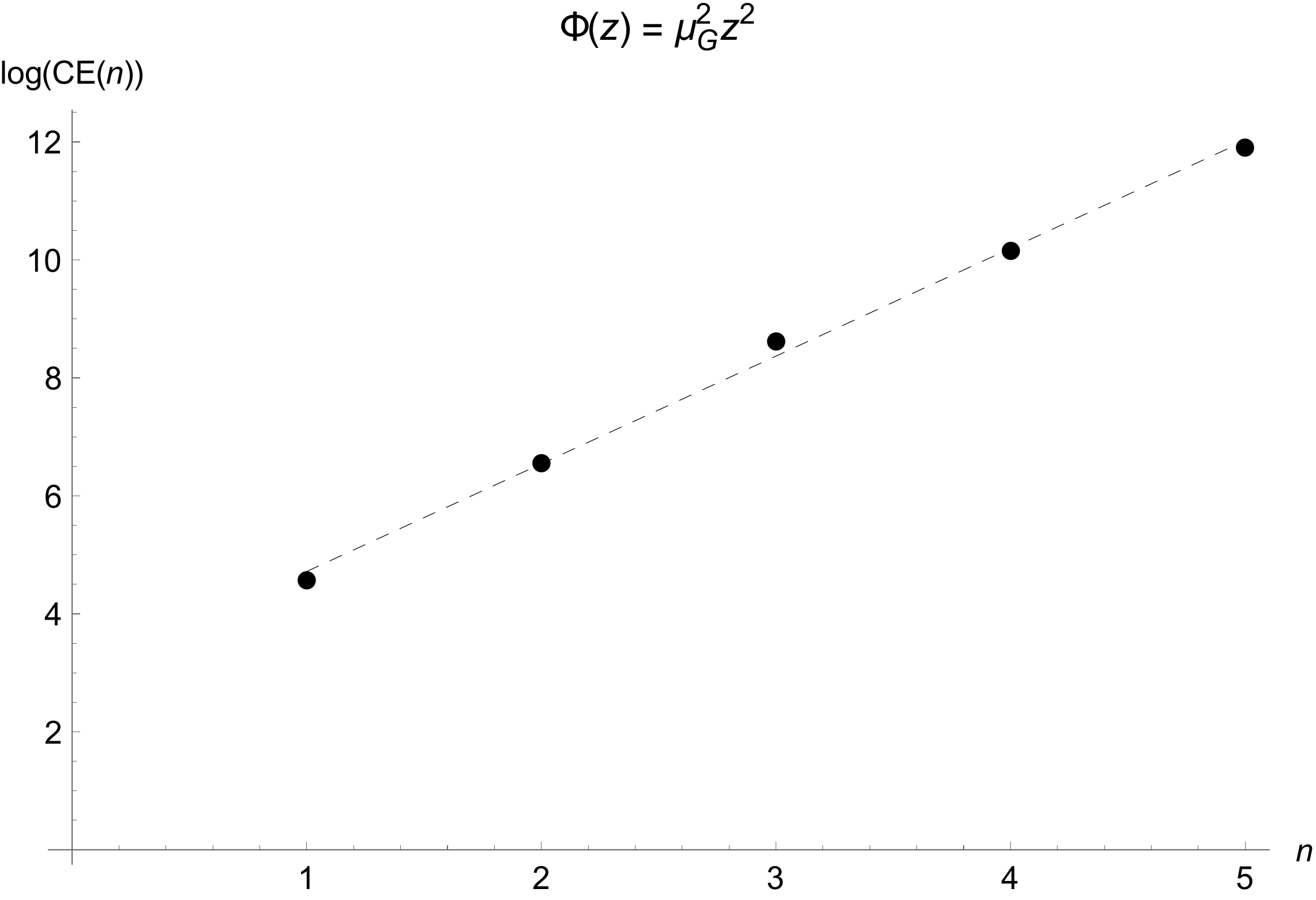}
\caption{CE of the pion family as a function of the $n$  quantum number, in the   quadratic dilaton AdS/QCD soft wall.}
\label{kaon}
\end{figure}

\begin{figure}[H]
\centering
\includegraphics[width=8.3cm]{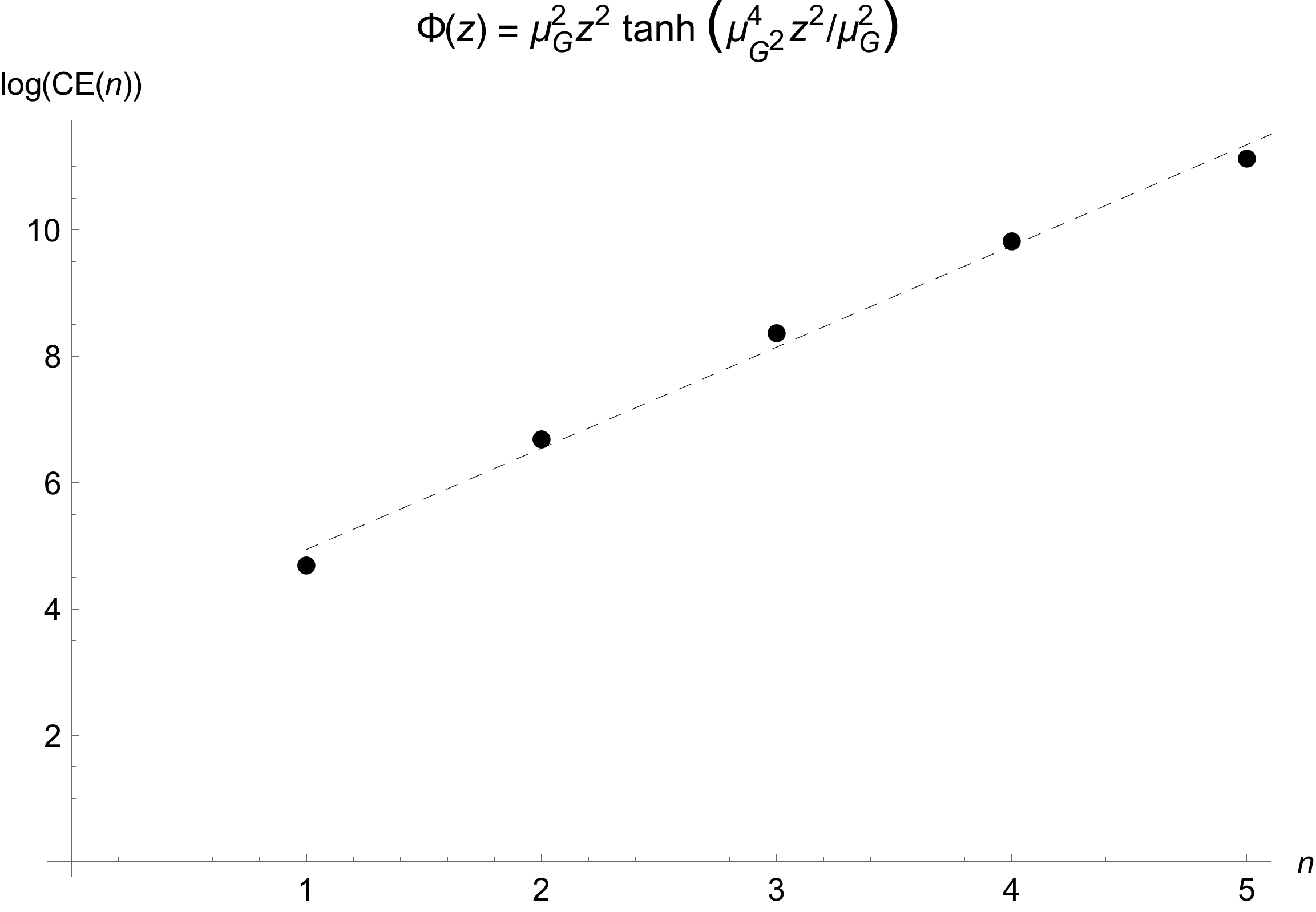}
\caption{CE of the pion family as a function of the $n$  quantum number, in the deformed dilaton AdS/QCD soft wall.} 
\label{f3}
\end{figure}

As the CE encodes how information measures  the shape  complexity of a physical system \cite{Gleiser:2018kbq,Gleiser:2014ipa,Sowinski:2015cfa}, computing the CE underlying the pion family requires its  energy density, $\upepsilon({\rm z})$.
This can be derived from a Lagrangian density, $\mathbb{L}$, with stress-momentum tensor components given by  
 \begin{equation}
 \!\!\!\!\!\!\!\!T_{MN}\!=\!  \frac{2}{\sqrt{ - g }}\!\! \left[ \frac{\partial (\sqrt{-g} \mathbb{L})}{\partial g^{MN} }\!-\!\partial_{ {\rm x}^Q }  \frac{\partial (\sqrt{-g}  \mathbb{L})}{\partial \left(\!\frac{\partial g^{MN} }{\partial {\rm x}^R}\!\right) }
  \right].
  \label{em1}
 \end{equation} 
 \noindent  
Hence, the energy density is read off the  $T_{00}({\rm z})$ tensor component in Eq. (\ref{em1}). 
The Fourier transform 
$\upepsilon(k) = \int_\mathbb{R}\upepsilon({\rm z})e^{-ik {\rm z}}\,d{\rm z},$ with respect to the ${\rm z}=1/\Lambda_{\rm QCD}$ dimension that defines the $\Lambda_{\rm QCD}$ energy scale, defines the so-called modal fraction    
\cite{Gleiser:2012tu,Sowinski:2015cfa}
\begin{eqnarray}
\upepsilon_\bullet(k) = \frac{|\upepsilon(k)|^{2}}{\int_{\mathbb{R}}  |\upepsilon(k)|^{2}dk}.\label{modalf}
\end{eqnarray} It consists of a probability distribution of correlation,  measuring the weight of a given  $k$ Fourier wave mode into the power spectrum \cite{Braga:2018fyc}. Hence, the CE is responsible to measure the amount of information that is encrypted into the spatial portrait of  the energy density. The CE reads \cite{Gleiser:2011di,Gleiser:2012tu,Gleiser:2013mga,Gleiser:2015rwa}
\begin{eqnarray}
S[\upepsilon] = - \int_{\mathbb{R}}{\widetilde{\upepsilon_\bullet}}(k)\log {\widetilde{\upepsilon_\bullet}}(k)\, dk\,,
\label{confige}
\end{eqnarray}
for $\widetilde{\upepsilon_\bullet}(k)=\upepsilon_\bullet(k)/{\upepsilon_{\bullet}}_{\rm max}(k)$.
Following this protocol, the CE can be forthwith computed for  the pseudoscalar pion meson family, first as a function of $n$ excitation number. Table \ref{cepion} shows the CE for both dilatonic field backgrounds (\ref{quadraticd}, \ref{tanh}). 
\begin{table}[h]
\begin{center}\medbreak
\begin{tabular}{||cc||c||c||c||}
\hline\hline
  &   $n$ & pion~CE  $(\upphi_1({\rm z}))$& pion~CE  $(\upphi_2({\rm z}))$\\ \hline\hline
     &  \, 1 \,&\, $97.5$&$77.9$ \\\hline
     \,&\,   2 \,&\, $702.1$&$506.3$ \\\hline
     \,&\,   3 \,&\, $5.613\times 10^3$&$3.129\times 10^3$\\\hline
     \,&   4\, &\, $2.587\times 10^4$&$1.548\times 10^4$ \\\hline
     \,&\,   5 \,&\, $1.490\times 10^5$&$1.275\times 10^5$ \\\hline
\hline\hline
\end{tabular}\caption{The CE for the pseudoscalar pion family, respectively in each column for the quadratic and deformed dilatonic field backgrounds. 
 }
\label{cepion}
\end{center}
\end{table}
 Table \ref{cepion} shows the CE of the pseudoscalar pion family, identifying the ${\mathtt{\pi}}_n$ eigenfunctions  in Eq. (\ref{motionphi}), as ${\mathtt{\pi}}_1=\{{\mathtt{\pi}}_\pm,  
{\mathtt{\pi}}_0\}$, ${\mathtt{\pi}}_2=\pi(1300)$, ${\mathtt{\pi}}_3=\pi(1800)$, ${\mathtt{\pi}}_4=\pi(2070)$, ${\mathtt{\pi}}_5=\pi(2360).$
Then, the  CE of each $\pi_n$ meson mode in the pion meson family is represented by the points in Fig. \ref{kaon}, strictly for the 
$\pi$ mesons in PDG. In addition, introducing linear regression yields the configurational entropic Regge trajectory, for the pion family, in the dashed line in Fig. \ref{kaon}.
\noindent Their explicit expression of the configurational entropic Regge trajectory reads
\begin{eqnarray}
 \log({\rm CE}_\pi(n)) &=& 1.8256\,n + 2.8927,\label{itp1}
   \end{eqnarray} within $\sim1\%$ standard deviation.
Similarly, the CE can be also computed, now taking into account the $\upphi_2$ dilaton (\ref{tanh}), illustrated in the third column of Table \ref{cepion}. Fig. \ref{f3} depicts, for the $\pi_n$ pion modes in PDG ($n\leq 5$), the computed results based in Table \ref{cepion}, illustrating the configurational entropic Regge trajectory in the deformed dilaton AdS/QCD soft wall. 
\noindent Linear regression yields the following relation between the configurational entropy of the pion family and the $n$ excitation number,  
\textcolor{black}{\begin{eqnarray}\label{itp2}
 \log({\rm CE}_\pi(n)) &=& 1.6015\,n + 3.3398, \end{eqnarray} within $\sim0.2\%$ standard deviation.}

A  scaling law can be then noticeable from Eqs. (\ref{itp1}, \ref{itp2}). It is implemented by the configurational entropic  
Regge trajectory, that in this case associates the CE with the $n$ excitation number, for the $\pi$ meson family. 
Motivated by the standard Regge trajectory in Fig. \ref{data1}, that relates the $\pi$ mesons squared mass spectra to the $n$ excitation number,
one can take the reverse way to predict, thus, the mass spectra of the further members in the $\pi$ meson family, that have been not yet 
experimentally observed. Indeed, computing the CE for the meson family with respect to the $\pi_n$ meson family mass, once the respective $n$ excitation is fixed, yields data that can be interpolated to  output the mass of the further members in the $\pi$ meson family, that have been not yet 
experimentally detected. 

\begin{figure}[H]
\centering
\includegraphics[width=8.3cm]{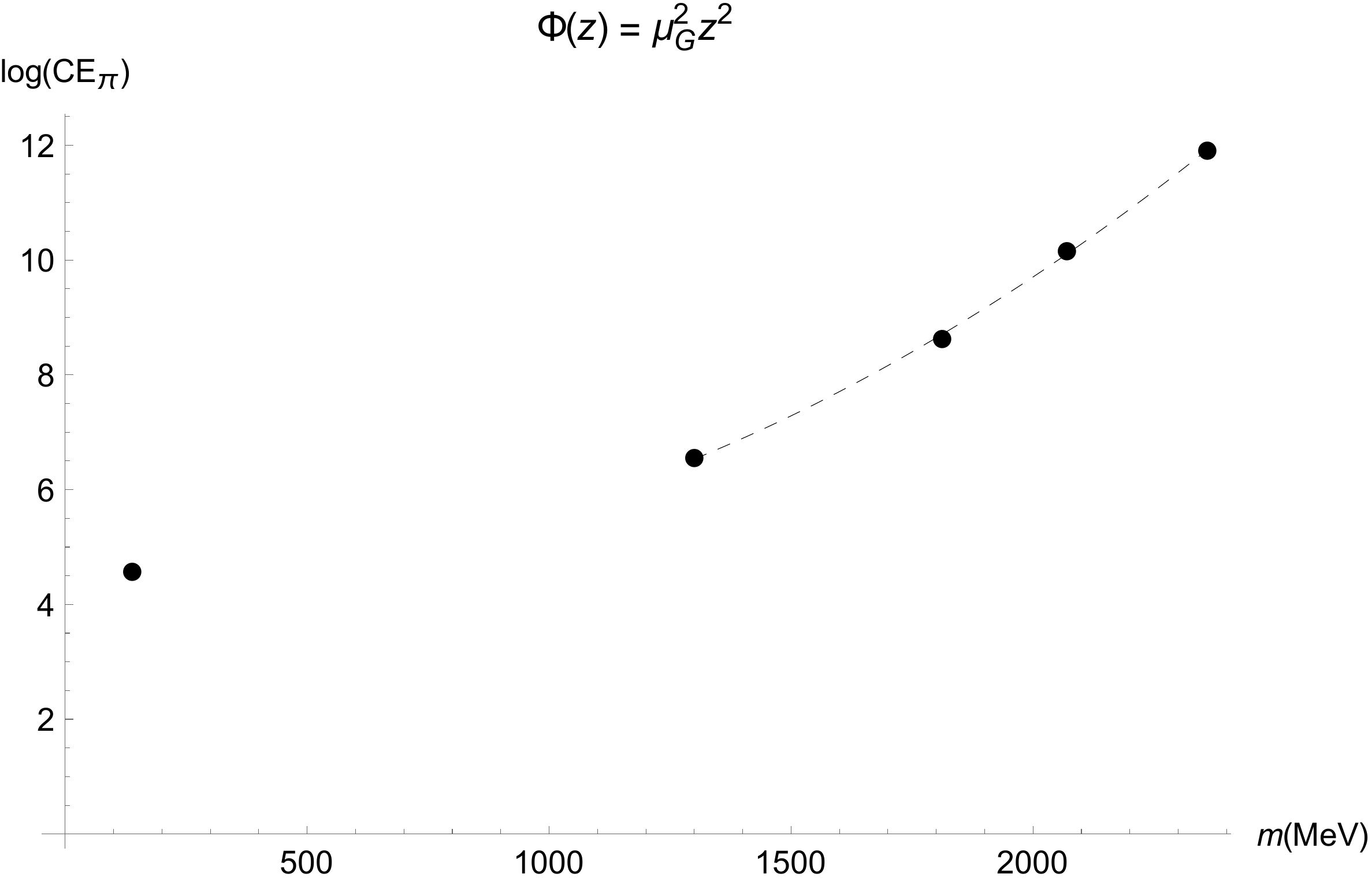}\medbreak
\includegraphics[width=8.3cm]{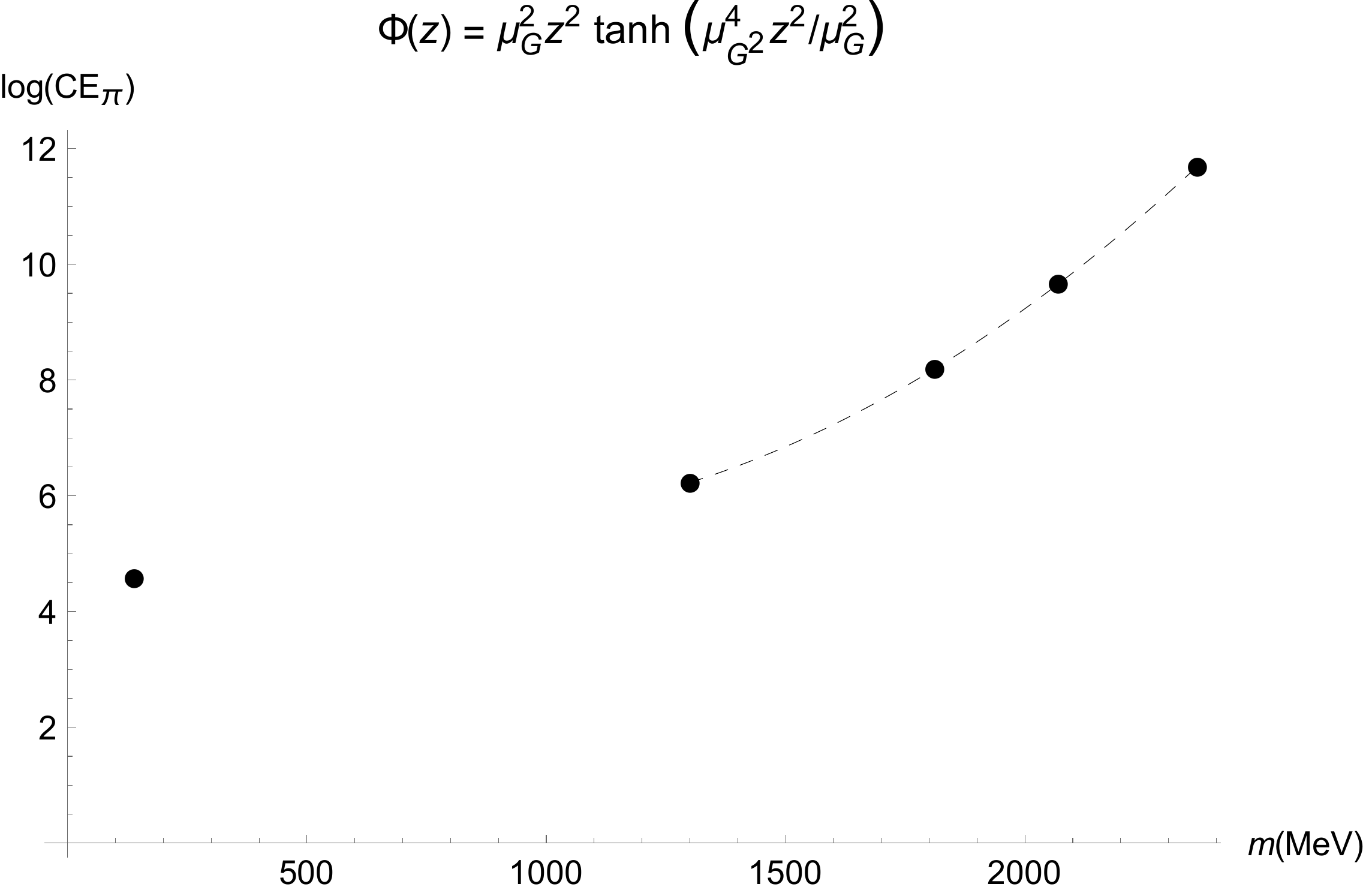}\medbreak
\caption{Configurational entropic Regge trajectory of the $\pi$ meson family with respect to the mass spectra, using the quadratic dilaton  (top panel) and deformed dilaton (bottom panel).}\label{f6}
\end{figure}

Firstly, the quadratic dilaton  (\ref{quadraticd}) shall be used to infer the mass spectra of the $\pi_n$ meson family, in the top panel in Fig. \ref{f6}. The  deformed dilatonic field (\ref{tanh}) is a refined model  that implements the mass spectra of 
the $\pi$ meson family, in great agreement to experimental data in PDG, having the  quadratic dilaton (\ref{quadraticd}) as the UV limit.
 The configurational entropic Regge trajectory, with respect to the $\pi$ meson family mass, is then shown in the bottom panel in Fig. \ref{f6}. 
\noindent Respectively for the plots in Fig.  \ref{f6}, the configurational entropic Regge trajectories have the following 
explicit expressions:
\begin{eqnarray}\label{itq1}
\!\!\!\!\!\!\!\! \log({\rm CE}_\pi(m)) \!=\!1.5337\!\! \times\!\! 10^{-6}
   m^2\!-\! 0.0005\, m\!+\!4.6301, \\
\!\!\!\!\!\!\!\! \log({\rm CE}_{\pi}(m)) \!=\!2.3869\!\!\times\!\! 10^{-6}
   m^2\!-\! 0.0036\,m\!+\!5.8556,\label{itq2}
   \end{eqnarray} within $\sim0.2\%$ standard deviations.
 Alternatively from calculating the mass spectra of the $\pi$ meson family, resolving the system  (\ref{motionpi},\ref{motionphi}), 
 the linear regression (\ref{itp1}, \ref{itp2}) and the quadratic interpolation (\ref{itq1}, \ref{itq2}), can be then employed altogether, 
 to calculate the CE of the $\pi$ meson family, for the $\pi_n$ excitations with $n> 5$, with good accuracy 
 at least for the first members, $\pi_n$ of the pion meson family after the  
 $\pi_5$ element. 
%

Let one denotes by $m_{\pi,n}$ the associated mass spectrum of the $n^{\rm th}$ pion,  in the pseudoscalar $\pi$ meson  family,   For the quadratic dilaton,  the mass spectra of the ${\mathtt{\pi}}_6$, ${\mathtt{\pi}}_7$  and ${\mathtt{\pi}}_{8}$ members of the $\pi$ meson family can be then inferred. In fact, for $n=6$, Eq. (\ref{itp1}) yields 
$\log({\rm CE}_\pi) = 13.846$.  Replacing this value in the configurational entropic Regge trajectory (\ref{itq1}), one obtains the mass $m_{\pi,6}=2630$ MeV, for the ${\mathtt{\pi}}_6$ mesonic state, yielding the reliable range $2612\; {\rm MeV}\lesssim m_{\pi,6}\lesssim 2648\;{\rm MeV}$, for the ${\mathtt{\pi}}_6$ pseudoscalar meson state. Analogously, the ${\mathtt{\pi}}_{7}$ meson has derived  mass $m_{\pi,7}=2861$ MeV. The standard deviations of Eqs. (\ref{itp1}) and (\ref{itq1}) then yield the reliable mass range $2839\lesssim m_{\pi,7}\lesssim 2883$ MeV. Once more,  the ${\mathtt{\pi}}_8$ pseudoscalar state has mass $m_{\pi,8}=3074$ MeV, and experimental standard deviations yield the  trustworthy range $3049\lesssim m_{\pi,7}\lesssim 3099$ MeV. 
On the other hand, for the deformed dilaton, considering  $n=6$ in Eq. (\ref{itp1}) implies that  
$\log({\rm CE}_\pi) = 12.948$.  Substituting it into the configurational entropic Regge trajectory (\ref{itq1}) yields the mass $m_{\pi,6}=2631$ MeV, within the experimental-based range $2612\; {\rm MeV}\lesssim m_{\pi,6}\lesssim 2649\;{\rm MeV}$, for the ${\mathtt{\pi}}_6$ pseudoscalar meson. Analogously, one can derive for the ${\mathtt{\pi}}_{7}$ mesonic state the mass $m_{\pi,7}=2801$ MeV, within the  reliable range $2779\lesssim m_{\pi,7}\lesssim 2823$ MeV, considering standard deviations in Eqs. (\ref{itp1}) and (\ref{itq1}). In addition,  the ${\mathtt{\pi}}_8$ meson presents mass equal to $m_{\pi,8}=2959$ MeV, establishing the phenomenologically reliable range $2934\lesssim m_{\pi,8}\lesssim 2984$ MeV for its mass. 
\section{Concluding remarks and perspectives}
\label{iv}
The main feature that distinguishes the $\pi$ meson family description in AdS/QCD, compared with  other light-flavour meson families, relies on the EOMs that govern the mesonic states modes and their mass spectra, for each meson family.
In fact, delving into the pions one can realize that they are the only family along with the $\phi$ mesons, among the $a_1$, $f_0$, and $\rho$ mesons in AdS/QCD, whose EOM \eqref{motionpi} has a potential that is a function of other meson, the $\phi$ meson. In this case,
there is a coupling between the EOMs of the $\pi$ and the $\phi$ mesons. This reflects a non-linear configurational entropic Regge trajectory for the pion family, with respect to the $\pi$ meson family mass, as shown in the plots in Fig. \ref{f6}, as well as the analytical expressions of the CE for these plots, in Eqs. (\ref{itq1}, \ref{itq2}). In addition, the configurational entropic Regge trajectory,  for the pion family, is linear with respect to the $n$ excitation number, as shown in the plots in Fig.  \ref{f6}, with explicit analytical expressions in Eqs. (\ref{itp1}, \ref{itp2}). It is a common feature with the $f_0$, $a_1$ and $\rho$ meson families studied in Ref.  \cite{Bernardini:2018uuy}. However, for these meson families
the  configurational entropic Regge trajectories, as a function of the respective meson families mass spectra, are also linear, contrary to the case here analyzed and depicted in the plots in Fig. \ref{f6}. 
It is then worth to emphasize that the reason for the non-linearity of  the configurational entropic Regge trajectory for the pion family, with respect to the $\pi$ meson mass spectra, is the coupling between the EOMs (\ref{motionpi}, \ref{motionphi}), with respective coupled potentials (\ref{v12}, \ref{v11}).

Another information provided by the configurational entropic Regge trajectories  is the values of the masses of the next generation of the $\pi$ states. Using the value CE of the $n^{\rm th}$ excitation, Eqs. (\ref{itp1}, \ref{itp2}), one can  employ Eqs. (\ref{itp1}) and (\ref{itq1}) to infer the mass spectra of the $\pi_{6}$, $\pi_{7}$  and  $\pi_{8}$, as discussed throughtout Sect. \ref{iii}. In the case of the quadratic dilaton the results found  are $m_{\pi,6}=2630 \pm 18$ MeV, $m_{\pi,7}=2861 \pm  22 $ MeV and $m_{\pi,8}=3074 \pm 25$ MeV. On the other hand, for the deformed dilaton the masses found are $m_{\pi,6}=2631 \pm 18$ MeV, $m_{\pi,7}=2801 \pm 22$ MeV and $m_{\pi,8}=2959 \pm 25$ MeV.  It is possible to improve these values of the masses with the eventual detection of  the pion excitation states, that shall contribute with more experimental points in Fig.  (\ref{data1}).

\medbreak
\paragraph*{Acknowledgments:}   L. F.  is supported by the National Council for Scientific and Technological Development  -- CNPq (Brazil) under Grant No. 153337/2018-4. RdR~is grateful to FAPESP (Grant No.  2017/18897-8) and to  CNPq (Grant No. 303293/2015-2), for partial financial support. 

\end{document}